\documentclass[%
 reprint,
nofootinbib,
 amsmath,amssymb,
 aps,
superscriptaddress
]{revtex4}

\usepackage{graphicx}
\usepackage{amsmath}
\usepackage{amssymb}
\usepackage{dcolumn}
\usepackage{bm}
\usepackage{color}
\usepackage{dsfont}
\usepackage{feynmp}
\usepackage{marvosym}
\usepackage{phaistos}

\usepackage{ulem}
\usepackage{cancel}

\newcommand \beq{\begin{eqnarray}}
\newcommand \eeq{\end{eqnarray}}

\definecolor{orange}{rgb}{.9,0.5,0.}
\definecolor{violet}{rgb}{.7,0.4,0.9}
\definecolor{vert}{rgb}{.3,0.7,0.}
\definecolor{bleu}{rgb}{.4,.7,1}
\definecolor{rouge}{rgb}{.8,0.2,0.2}

\begin{document}
\unitlength=1mm
\allowdisplaybreaks

\title{A novel background field approach to 

the confinement-deconfinement transition}

\author{Duifje Maria van Egmond}
\affiliation{Centre de Physique Th\'eorique, CNRS, Ecole polytechnique, IP Paris, F-91128 Palaiseau, France.}

\author{Urko Reinosa}
\affiliation{Centre de Physique Th\'eorique, CNRS, Ecole polytechnique, IP Paris, F-91128 Palaiseau, France.}

\author{Julien Serreau}
\affiliation{Universit\'e de Paris, CNRS, Astroparticule et Cosmologie, F-75013 Paris, France.}

\author{Matthieu Tissier}
\affiliation{Laboratoire de Physique Th\'eorique de la 
Mati\`ere Condens\'ee, UPMC, CNRS UMR 7600, Sorbonne
Universit\'es, 4 Place Jussieu,75252 Paris Cedex 05, France.}

\date{\today}

\begin{abstract}
We propose a novel approach to the confinement-deconfinement transition 
in Yang-Mills theories in the context of gauge-fixed calculations. The method is based on a background-field generalisation of the Landau gauge (to which it reduces at vanishing temperature) with a given, center-symmetric background. This is to be contrasted with most implementations of background field methods in gauge theories, where one uses a variable, self-consistent background. Our proposal is a bona fide gauge fixing that can easily be implemented on the lattice and in  continuum approaches. The resulting gauge-fixed action explicitly exhibits the center symmetry of the nonzero temperature theory that controls the confinement-deconfinement transition. We show that, in that gauge, the electric susceptibility diverges at a second order transition [{\it e.g.}, in the SU(2) theory], so that the gluon propagator is a clear probe of the transition. We implement our proposal in the perturbative Curci-Ferrari model, known for its successful description of various infrared aspects of Yang-Mills theories, including the confinement-deconfinement transition. Our one-loop calculation confirms our general expectation for the susceptibility while providing transition temperatures in excellent agreement with the SU(2) and SU(3) lattice values. Finally, the Polyakov loops above the transition show a more moderate rise, in contrast to previous implementations of the Curci-Ferrari model using a self-consistent background, and our SU(3) result agrees quite well with lattice results in the range $[0,2T_c]$.
\end{abstract}

\maketitle

\section{Introduction}

Quantum chromodynamics, the modern theory of the strong interaction, predicts a dramatic change in the behavior of thermodynamics observables at temperatures of the order of a hundred of MeV \cite{Borsanyi:2013bia}. This signals a qualitative change in the nature of the relevant degrees of freedom, from hadronic ones at low temperatures, to what is called a quark-gluon plasma at high temperatures. In the limit of infinitely heavy quarks, where the pure non-Abelian gauge dynamics is described by a Yang-Mills (YM) theory, this takes the form of an actual phase transition between phases with either confined or unconfined static quarks, governed by the spontaneous breaking of a symmetry peculiar to the nonzero temperature problem, the center symmetry \cite{Svetitsky:1985ye,Pisarski:2002ji}. 

This confinement-deconfinement transition in YM theories has been firmly established by means of lattice calculations \cite{Gavai:1982er,Gavai:1983av,Celik:1983wz} and serves as a stringent test for the various approximations involved in continuum approaches. In this latter case, an additional difficulty comes from the obligation to work in a gauge-fixed setting, which may not inherit the center symmetry of the YM action. Such an explicit breaking of the center symmetry by the gauge-fixing procedure should not alter the physical results in principle, but this is clearly an issue when approximations are involved. That is, for instance, the case in the widely used Landau gauge, where the lack of explicit center symmetry leads to unphysical results in approximate continuum calculations, as compared to the results of gauge-fixed lattice simulations \cite{Fister:2011uw,Reinosa:2013twa}. 

One important physical ingredient in the study of the phase transition is the presence of a nontrivial order parameter for the center symmetry, given, {\it e.g.}, by the Polyakov loop \cite{Polyakov:1978vu}. The latter involves nonperturbatively large field configurations in the (Euclidean) temporal direction $A_0\sim 1/g$, where $g$ is the coupling  constant. One convenient way to cope with the issue of properly encoding both the center symmetry of the problem and the large field configurations associated to the order parameter is to work with background field gauge methods \cite{Abbott:1980hw,Abbott:1981ke,KorthalsAltes:1993ca,Braun:2007bx}. This involves an arbitrary background field configuration $\bar A$ which enters the gauge-fixing condition and which, roughly speaking, acts as a sort of source term which can be chosen to favor the desired field configurations $A$ under the (gauge-fixed) path integral. In practice, working with two gauge fields is cumbersome and it is simpler to work with what are called self-consistent backgrounds, which are tuned to equal the field average, $\bar A=\langle A\rangle$. This has, indeed, revealed very efficient to correctly describe the confinement-deconfinement phase transition in YM theories in a variety of nonperturbative \cite{Braun:2007bx,Marhauser:2008fz,Braun:2009gm,Fischer:2013eca,Fischer:2014vxa,Quandt:2016ykm} and perturbative \cite{Reinosa:2014ooa,Reinosa:2015gxn,Reinosa:2015oua,Maelger:2017amh,Shibata:2015ywl} continuum approaches.

However, despite its many interesting advantages, this approach also comes with some drawbacks. The main one is that it is somewhat formal because it relies on using the background field both as a gauge-fixing tool and as a dynamical variable. In particular, this makes it difficult to implement in---and thus to compare to---lattice calculations. For instance, an interesting question, put forward in the Landau gauge \cite{Heller95,Heller97,Cucchieri00,Cucchieri01,Cucchieri07,Aouane:2011fv,Fischer10,Cucchieri11,Cucchieri12,Silva:2013maa}, is whether the phase transition is directly encoded in (gauge-fixed) propagators, in particular, in the electric component ({\it i.e.}, in the time direction) of the gluon propagator at vanishing momentum and frequency. Gauge-fixed lattice calculations in the Landau gauge show a strong increase of the electric susceptibility\footnote{We define the susceptibility as the zero-frequency and zero-momentum limit of the propagator. Our definition differs from that in \cite{Maas:2011ez}.} at the second order phase transition of the SU($2$) theory, which is not seen in continuum calculations \cite{Fister:2011uw,Reinosa:2013twa}, probably for the lack of proper account of the order parameter mentioned above. Although the phase transition is correctly described using the self-consistent background field approach, it is not clear how to compare to those lattice results in the Landau gauge. On the other hand, the gluon propagator has been computed for the SU($2$) theory in the presence of a self-consistent background \cite{Reinosa:2016iml} and one observes a clear signal of the phase transition. However, there exists, so far, no lattice results to compare to. 

In the present article, we propose a novel, simple approach to the confinement-deconfinement transition that exploits the advantages of both the Landau gauge and the background field gauges, while avoiding the drawbacks mentioned above. This is based on using a background field gauge with a particular, fixed (non-dynamical) background which is invariant under center transformations. This yields a bona-fide gauge-fixed theory that is, first, easily implementable in lattice calculations (with techniques currently employed for the Landau gauge, see for instance \cite{Cucchieri12,Silva:2013maa}) and, second, explicitly center symmetric, allowing one to keep track of this central aspect of the problem in any continuum approach even in the presence of approximations. Also, the resulting effective action encodes, at once, the phase transition---with the average gluon field $\langle A\rangle$ playing the role of the order parameter---and the vertex functions of the theory, allowing for a simple study of the imprints of the former on the latter, that is easily testable in lattice calculations. For instance, we show that the above-mentioned electric susceptibility trivially diverges at the transition in the SU($2$) theory. Finally, the proposed gauge fixing reduces to the Landau gauge at vanishing temperature and thus can be viewed as a simple generalization of the latter at nonzero temperature.

After describing the method, its properties and its lattice implementation, we explicitly apply the center-symmetric gauge fixing in the context of the perturbative Curci-Ferrari (CF) approach to infrared QCD \cite{Curci:1976bt,Tissier:2010ts,Tissier:2011ey,Pelaez:2021tpq}. The latter is motivated by two essential observations from lattice simulations in the Landau gauge, namely, the fact that the coupling (defined in the Taylor scheme) remains finite and under perturbative control at all scales and the fact that the gluon propagator at zero momentum is finite, corresponding to the dynamical generation of a screening mass \cite{Bogolubsky:2009dc,Cucchieri:2009zt,Duarte:2016iko}. The CF Lagrangian in the Landau gauge simply corresponds to adding an effective tree-level gluon mass to the standard Faddeev-Popov Lagrangian. It has been successfully used to compute various infrared properties of various YM and QCD-like theories in the vacuum and at nonzero temperature and density \cite{Tissier:2010ts,Tissier:2011ey,Pelaez:2013cpa,Gracey:2019xom,Barrios:2020ubx,Reinosa:2014ooa,Reinosa:2014zta,Reinosa:2015oua,Reinosa:2020mnx}. In the latter case, it has been implemented in the framework of self-consistent background field techniques mentioned above and gives remarkable results already at one-loop order. In YM theories, it correctly predicts the order of the transition with transition temperatures in good agreement with known values from lattice calculations. Here, we compute the effective potential for constant temporal gauge fields in the proposed center-symmetric gauge at one-loop order. We find that, again, this correctly captures the phase structure of YM theories and we find values of the transition temperatures for SU($2$) and SU($3$) in remarkable agreement with lattice values, better than in the previous (self-consistent) background field formulation. Our results confirm that the SU(2) electric susceptibility diverges at the transition. In the SU(3) case, the singularity is replaced by a sharp peak at the transition, which could still be identifiable in lattice simulations. We also compute the Polyakov loop as a function of the temperature, which shows various improvements over a similar evaluation in the CF model using self-consistent backgrounds \cite{Reinosa:2014ooa}, for instance, a slower rise above the transition, in line with lattice results \cite{Kaczmarek:2002mc,Gupta:2007ax,Lo:2013hla,Dumitru:2012fw} as well as with results obtained within the functional renormalization group \cite{Herbst:2015ona}.

\section{The confinement-deconfinement transition and background fields}

\subsection{Finite temperature and center symmetry}
At nonzero temperature, YM theories are usually formulated in terms of the Euclidean action
\beq\label{eq:YM}
\int_x \frac{1}{4}F_{\mu\nu}^a(x)F_{\mu\nu}^a(x)\,,
\eeq
with $\smash{F_{\mu\nu}^a\equiv\partial_\mu A_\nu^a-\partial_\nu A_\mu^a+gf^{abc}A_\mu^bA_\nu^c}$ the Euclidean field-strength tensor and $g$ the coupling constant. The space-time integration is taken over a compact time interval, $\int_x\equiv \int_0^\beta d\tau \int d\vec{x}$, with $\smash{\beta\equiv 1/T}$ the inverse temperature, and the gauge field is periodic with period $\beta$: $A_\mu^a(\tau+\beta,\vec{x})=A_\mu^a(\tau,\vec{x})$. Among the gauge transformations\footnote{For a field $X^a$ in the algebra, we use the notation $\smash{X\equiv X^a t^a}$, where the $t^a$ are the generators of the gauge group.}
\beq\label{eq:U}
A^U_\mu(x)=U(x)A_\mu(x)U^\dagger(x)+\frac{i}{g}U(x)\partial_\mu U^\dagger(x)\,,
\eeq
that leave the YM action  invariant, only those that preserve the periodicity of the gauge field are actual symmetries of the theory.\footnote{To be qualified as an actual symmetry of the problem at finite temperature, the gauge transformations need to preserve not only the classical action but also the integration measure and the domain of integration under the functional integral.} They form a group, denoted ${\cal G}$ in what follows, which is actually larger than the group ${\cal G}_0$ of $\beta$-periodic gauge transformations that leaves physical observables invariant. It is easily shown that ${\cal G}$ is the group of gauge transformations that are $\beta$-periodic up to an element $W$ of the center of the gauge group: $\smash{U(\tau+\beta,\vec{x})=WU(\tau,\vec{x})}$ \cite{Svetitsky:1985ye,Pisarski:2002ji}. For instance, in the case of SU($N$), the center elements are of the form $\smash{W=w\mathds{1}}$, with $w\in\left\{e^{i2\pi k/N};k=0,\dots,N-1\right\}$. The finite temperature theory thus possesses an enlarged physical symmetry described by the quotient group ${\cal G}/{\cal G}_0$.\footnote{That ${\cal G}/{\cal G}_0$ is a group
  follows from the fact that ${\cal G}_0$ is a normal subgroup within
  ${\cal G}$.} The latter is isomorphic to the center of the
gauge group and is known as the center-symmetry group
\cite{Svetitsky:1985ye,Pisarski:2002ji}.

There exist many order parameters for this symmetry, some of which
will be discussed in this article. The most popular one is the Polyakov
loop [here for SU($N$)]
\beq \ell\equiv\frac{1}{N}\left\langle{\rm tr}\,{\cal
    P}\,\exp\left\{ig\int_0^\beta
    d\tau\,A_0^a(\tau,\vec{x})t^a\right\}\right\rangle,\label{eq:ell}
\eeq 
which, being invariant under ${\cal G}_0$ and, thus, a physical observable, is directly measurable in lattice simulations.
Under any transformation $\smash{U\in {\cal G}}$, $\ell$ is multiplied by
the corresponding center phase $w$, which implies that it vanishes if the symmetry
group ${\cal G}/{\cal G}_0$ is unbroken and that a nonzero value signals the spontaneous breaking of the center symmetry. The relevance of these
considerations for the deconfinement transition is that the Polyakov
loop grants access to the free-energy $F_q$ of a static quark source
$\ell\propto e^{-\beta F_q}$ \cite{Polyakov:1978vu}. The phase of
unbroken center symmetry, with $\smash{\ell=0}$, corresponds then to a
phase where such a static quark source would have an infinite free energy, that is the
confined phase of the system. Phrased differently, studying the
spontaneous breaking of center symmetry gives access to the
deconfinement transition. When trying to export this
  discussion within a practical, and necessarily approximated, computational scheme in the continuum,
  one faces the problem that typical gauge-fixed actions break explicitly 
the center symmetry, which obviously hinders the analysis of
   its spontaneous breaking. As we now recall,
  one way out is to upgrade the standard gauge fixing into a
  background gauge fixing.

\subsection{The class of Landau-DeWitt gauges}
The following discussion applies in principle to any background field gauge. For definiteness, however, we consider the background Landau (or Landau-DeWitt) gauges, defined by the condition 
\beq
\bar D_\mu(A_\mu^a-\bar A_\mu^a)=0\,,\label{eq:condition}
\eeq 
where $\bar A_\mu^a$ is a given background field configuration and $\smash{\bar D_\mu^{ab}\equiv \partial_\mu\delta^{ab}+gf^{acb}\bar A_\mu^c}$ is the associated adjoint covariant derivative. The condition (\ref{eq:condition}) should be seen as defining a family of gauges parametrized by a field $\bar A_\mu^a$ of gauge fixing parameters. Choosing the background in one way or another defines a specific gauge in this family. In this work, we consider two particular choices that are convenient at finite temperature.

The relevance of background gauges lies in that, provided one allows the background $\bar A_\mu^a$ to be gauge-transformed, the gauge condition (\ref{eq:condition}) transforms covariantly. It is then possible, in a certain sense, to export the gauge-invariance of the YM theory (\ref{eq:YM}) 
into the gauge-fixed setting. For instance, introducing the usual ghost, antighost, and Nakanishi-Lautrup fields, $c$, $\bar c$, and $h$, respectively, the Faddeev-Popov action 
\beq\label{eq:action}
S[A,\bar A]=\int_x\left\{\frac{1}{4}F_{\mu\nu}^aF_{\mu\nu}^a+\bar D_\mu\bar c^a  D_\mu c^a+ih^a\bar D_\mu(A_\mu-\bar A_\mu^a)\right\},
\eeq
associated to the gauge-fixing condition (\ref{eq:condition}) is invariant under the simultaneous gauge transformation (\ref{eq:U}) of the gauge field $A_\mu^a$ and the background $\bar A_\mu^a$ (and the color rotation $\varphi^U=U\varphi U^\dagger$ of the other fields $\varphi=c,\bar c,h$ that we have left implicit):
\beq\label{eq:sym}
S[A^U,\bar A^U]=S[A,\bar A]\,.
\eeq
This property is of utmost importance at nonzero temperature for it encodes the center symmetry.


In continuum approaches, one efficient way to study the spontaneous breaking of a symmetry is through the effective action. In background field gauges, this is a functional  $\Gamma[A,\bar A]$ of both the (average)\footnote{Using the standard convention, we denote with the same letter the fluctuating field that apperas as an argument of the classical action under the functional integral and the average field that appears as an argument of the effective action.}  field $A_\mu^a$ and the background $\bar A_\mu^a$. The linearly realized symmetry (\ref{eq:sym}) trivially implies
\beq
\Gamma[A^U,\bar A^U]=\Gamma[A,\bar A]\,,\label{eq:sym2}
\eeq
for any $U\in {\cal G}$ \cite{Braun:2007bx,Reinosa:2015gxn,Reinosa:2020mnx}. This identity does not yet provide a full grasp on the center symmetry, however, for it does not allow to discriminate between center-symmetric and center-breaking states. Indeed, for a given background $\bar A$, the state of the system is obtained as the minimum $A_{\mbox{\tiny min}}[\bar A]$ of $\Gamma[A,\bar A]$ with respect to $A$. The symmetry identity (\ref{eq:sym2}) implies that, under any transformation $\smash{U\in {\cal G}}$, one has $\smash{A^U_{\mbox{\tiny min}}[\bar A]=A_{\mbox{\tiny min}}[\bar A^U]}$. Thus, as the state is transformed, the background that is chosen for the description of the states (and therefore the gauge fixing itself) is changed as well. This makes it difficult to identify the center-symmetric states and thus to decide whether or not the symmetry is spontaneously broken. 

The (by now standard) proposal of Ref.~\cite{Braun:2007bx} to fully account for center symmetry within background field gauges relies on the use of the {\it background effective action}, defined as $\smash{\tilde\Gamma[\bar A]\equiv\Gamma[A=\bar A,\bar A]}$. It follows from Eq.~(\ref{eq:sym2}) that the latter is invariant under the gauge transformation of its argument, $\smash{\tilde\Gamma[\bar A^U]=\tilde\Gamma[\bar A]}$, for any $\smash{U\in {\cal G}}$, which, in particular, includes the center symmetry. It follows that the minima of $\tilde \Gamma[\bar A]$---which can be shown to represent the possible states of the system\footnote{Note that is not a trivial statement because $\tilde \Gamma[\bar A]$ is not an effective action in the usual sense. In particular, it is not the Legendre transform of the generating functional for connected correlators.}---are alternative order parameters for the confinement-deconfinement transition \cite{Reinosa:2015gxn,Herbst:2015ona,Reinosa:2020mnx}. The great benefit of this approach is that the background effective action can be easily evaluated for simple, constant background field configurations (see below), which allows for a convenient description of the phase transition. 

This also comes with some drawbacks though. The first one is that $\tilde\Gamma[\bar A]$ is a somewhat formal object in the sense that its argument, the background field, actually serves to define a gauge. Changing the argument thus amounts to changing the gauge. Because of that, gauge-fixed quantities (such as propagators and vertices) are not directly accessible from this functional. Moreover, this approach relies on implicit assumptions that are not trivially satisfied in practical calculations. One crucial such assumption is the independence of the partition function (a physical, gauge-invariant quantity) on the background field (a gauge-fixing parameter) \cite{Reinosa:2016iml,Reinosa:2020mnx}. Finally, a related point, is that the background effective action is not directly accessible in (gauge-fixed) lattice calculations. The proposal below aims at avoiding these drawbacks while keeping the main advantage of the background effective action approach, that is, the fact that, contrarily to, say, the Landau gauge, it conveniently encodes the center symmetry of the nonzero temperature theory.

\subsection{Center-symmetric effective action and effective potential}\label{sec:constant}

We aim at finding a gauge condition of the type \eqref{eq:condition} with a given, fixed background but with explicit center symmetry. This is easily realized by  choosing a {\it center-symmetric background}, that we now define precisely. As emphasized above, the actual center symmetry group is not ${\cal G}$ but rather ${\cal G}/{\cal G}_0$. In a certain sense, the transformations of ${\cal G}_0$ should be seen as true (unphysical) gauge transformations and the physical content of the symmetry group ${\cal G}$ is captured by ${\cal G}/{\cal G}_0$ once the redundancy associated to ${\cal G}_0$ has been quotiented away. Center-symmetric backgrounds are represented by configurations $\bar A_c$ invariant under ${\cal G}$ modulo ${\cal G}_0$:
\beq\label{eq:inv_orbit}
\forall\,U\in {\cal G},\quad\exists\,U_0[U]\in {\cal G}_0,\quad \bar A^U_c=\bar A^{U_0[U]}_c\,.
\eeq
They can be identified using the notion of {\it Weyl chambers} \cite{Herbst:2015ona,Reinosa:2015gxn,Reinosa:2020mnx}. Choosing such a background, we define a center-symmetric effective action as $\Gamma_c[A]\equiv\Gamma[A,\bar A_c]$. It is such that
\beq\label{def}
\Gamma_c[A]=\Gamma[A,\bar A_c]=\Gamma[A^U,\bar A^U_c]=\Gamma[A^U,\bar A^{U_0[U]}_c]=\Gamma[A^{U_0^{-1}[U]U},\bar A_c]=\Gamma_c[A^{U_c}]\,,
\eeq
with $\smash{U_c\equiv U_0^{-1}[U]U}$. Since $U_c$ represents the same center transformation as
$U$ in ${\cal G}/{\cal G}_0$, we deduce that $\Gamma_c[ A]$ is invariant under center transformations.\footnote{We stress here that $\Gamma_c[A]$ is not invariant under every transformation $U\in{\cal G}$ but, rather, under certain representatives $U_c$ of each of the transformations in ${\cal G}/{\cal G}_0$ which are the {\it physical} center transformations, where the gauge redundancy has been quotiented away. The important point for our present purposes is that each element of ${\cal G}/{\cal G}_0$ is represented and that there exists center-invariant $A$-configurations that define the confining configurations. These are provided by the configurations $A=\bar A_c$ which are invariant under $U_c$ owing to (\ref{eq:inv_orbit}).} In turn the minima $A_c^{\rm min}=A_{\mbox{\tiny min}}[\bar A_c]$ of $\Gamma_c[A]$ are order parameters for the center symmetry, just as the minima of $\tilde\Gamma[\bar A]$ or the Polyakov loop $\ell$. In summary, rather than studying the transition 
in the $A=\bar A$ subspace of the $(A,\bar A)$ space, we can alternatively use the $\bar A=\bar A_c$ subspace. 

One benefit of this approach is that it gives direct access to the propagator which is nothing but the inverse of $\delta^2\Gamma[A,\bar A_c]/\delta A_\mu^a(x)\delta A_\nu^b(y)$ evaluated for $\smash{A=A_c^{\rm min}}$. In particular, this tells us immediately that the inverse propagator should develop a zero mode at a continuous transition, {\it e.g.}, in the SU(2) theory. As we discuss in the next subsection, this is, in principle, easily testable in lattice simulations. In Sec.~\ref{sec:CF}, we provide evidence for this expectation in the context of the CF model.

To make the discussion more concrete, let us note that it is enough to restrict to constant temporal backgrounds in the diagonal---or commuting---part of the gauge group algebra, the Cartan subalgebra, that is,
\beq\label{eq:form}
\beta g\bar A_{\mu}=\bar r^j t^j T \delta_{\mu0}\,,
\eeq
with $[t^j,t^{j'}]=0$. The vector $\bar r$, of components $\bar r^j$, lives in the Cartan space, which is spanned by physically equivalent Weyl chambers, whose discrete symmetries actually represent the various symmetries of the problem. For instance, in the SU($2$) case, the Cartan subalgebra is one-dimensional, with the single generator $t=\sigma^3/2$, where $\sigma^i$ are the Pauli matrices, and an elementary Weyl chamber is the segment $[0,2\pi]$. A center transformation is described by $\bar r\to 2\pi-\bar r$ and the center-symmetric point is thus $\bar r_c=\pi$. This can be generalized to an arbitrary gauge group. 
We can thus choose the center-symmetric effective action, with a slight abuse of notation as compared to Eq.~\eqref{def},
\beq\label{eq:bckrc}
 \Gamma_c[A]=\Gamma[A,\bar r_c]\,,
\eeq
which is the generating functional of one-particle-irreducible vertex functions in the gauge\footnote{Note that the actual background field $\bar A_c$  entering the definition of the gauge in Eq.~\eqref{eq:condition} is proportional to the temperature; see Eq.~\eqref{eq:form}.} defined by $\bar r_c$. 

\begin{figure}[t]
  \centering
 \includegraphics[width=0.55\linewidth]{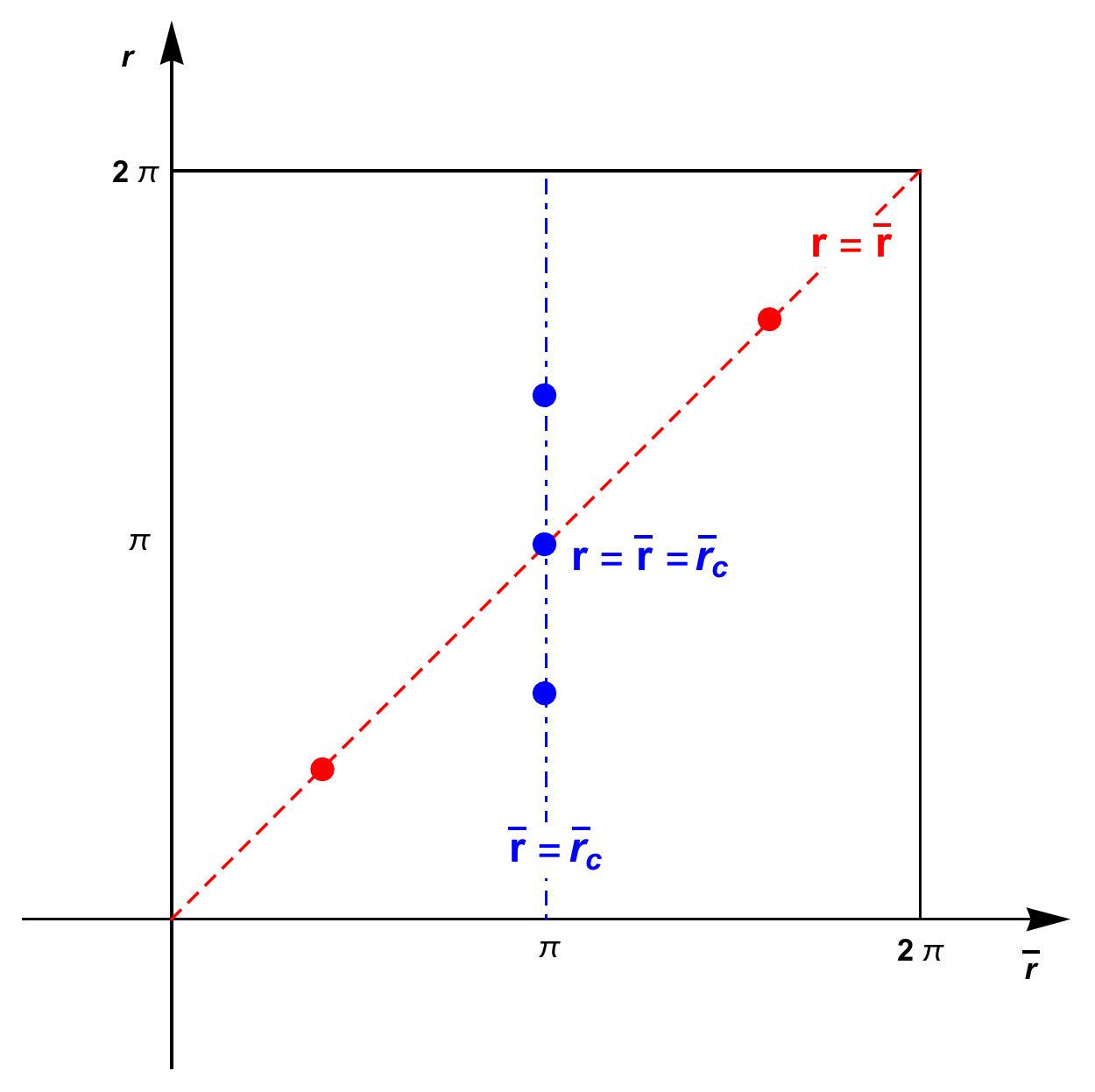}
 \caption{The two possible strategies to discuss the breaking of the center symmetry in background field gauges (here in the SU($2$) case for the purpose of illustration): the standard approach is based on the {\it background} potential $\tilde V(\bar r)$, defined along the (red) dashed line whereas the present proposal is based on the {\it center-symmetric} potential $V_c(r)$, defined along the (blue) dash-dotted line. Both functions have the center-symmetry in the form $x\to 2\pi-x$ with $x=r$ or $x=\bar r$, as illustrated by the two pairs of points connected by center symmetry.}\label{fig:strategies}
\end{figure}

For the purpose of studying the phase transition, we only need to find the minimum $A_c^{\rm min}$ of $\Gamma_c[A]$. With the choice of background \eqref{eq:form}, the latter is of the form  
\beq\label{eq:form2}
\beta g A_{\mu}=r^j t^j T \delta_{\mu0}\,,
\eeq
and we can further simplify the analysis by restricting to such configurations, for which the functional \eqref{eq:bckrc} reduces to a simple effective potential $\Gamma_c[A]\propto V_c(r)$. The latter is invariant under the center transformations of its argument. For instance, in the SU(2) case, we have 
\beq
V_c(2\pi- r)=V_c(r)\,.\label{eq:center_sym}
\eeq
Studying the dynamical breaking of the center symmetry and the possible phase transition simply amounts to finding the minima of $V_c(r)$. A symmetric state corresponds to $r_{\rm min}=\pi$ and any departure signals the spontaneous breaking of the center symmetry. In Fig.~\ref{fig:strategies}, using this simple SU(2) example, we give a graphical representation of the present proposal, using the potential $V_c(r)$, as compared to the usual one based on the background potential $\tilde V(\bar r)$, obtained from the background effective action defined above for a background of the form \eqref{eq:form}.

As mentioned before, one clear advantage of the present approach as compared to that based on the background effective action is that the phase transition is  described in terms of a gauge-fixed effective action (or potential) and is thus directly related to standard vertex functions. For instance, the vanishing of the second derivative $V''_c(r=\pi)$ at a second order phase transition in the SU($2$) theory (see the explicit calculation below) is imprinted in the (electric) gluon correlator at vanishing momentum and frequency, which is proportional to $1/V''_c(r_{\rm min})$, which diverges at the transition.

\subsection{Lattice implementation}\label{sec:lattice}
The center-symmetric gauge proposed in this work can easily be implemented on the lattice. To see this, we note that it simply amounts to the standard Landau gauge fixing with, however, twisted boundary conditions for the gauge fields.\footnote{ The use of twisted boundary conditions to explore the confinement-deconfinement phase transition is not new, see for instance Ref.~\cite{Bilgici:2008qy,Fischer:2009wc} where the notions of dressed Polyakov loops and dual condensates are introduced. Our proposal is here that the background Landau gauge with center-symmetric background can be easily simulated on the lattice by implementing different boundary conditions to existing routines for gauge-fixing in the Landau gauge.} Let us illustrate the point with the SU($2$) theory in the center-symmetric background gauge \eqref{eq:condition} with $g\bar A_{c,\mu}^a(x)=\pi T \delta_{\mu0} \delta^{a3}\sigma^3/2$. Consider the transformation
\beq\label{eq:tr}
U(\tau)=\exp\left(i\pi\tau T \frac{\sigma^3}{2}\right)\,,
\eeq
which does not belong to ${\cal G}$ since $U(\beta)=i\sigma_3 U(0)$ and is obviously not a symmetry of the theory. One easily checks that it transforms the background field to zero: $g\bar A^U_{c,\mu}(x)=0$. In that sense, the transformation \eqref{eq:tr} relates the center-symmetric gauge to the usual Landau gauge. To see how this affects the fluctuating gauge field under the path integral, let us consider the variable $\smash{a_\mu=A_\mu-\bar A_\mu}$ which, as one readily checks, transforms as $a^U_\mu(x)=U(x)a_\mu(x)U^\dagger(x)$ under a SU($2$) transformation. Now, decompose $a_\mu(x)=a_\mu^\kappa(x)t^\kappa$ in the basis $t^\kappa=\{t^0,t^\pm\}$, with $t^0\equiv\sigma^3/2$ and $t^\pm\equiv(\sigma^1\pm i\sigma^2)/2\sqrt{2}$. Using $[t^0,t^\pm]=\pm t^\pm$, one obtains 
that $[a^U_\mu]^\kappa(x)=e^{i\kappa\pi\tau T}a_\mu^\kappa(x)$ and, in particular, the field $[a^U_\mu]^0$ remains periodic in Euclidean time whereas the fields $[a^U_\mu]^\pm$ become anti-periodic.\footnote{For the continuum formulation, the same is true for the ghost and the Nakanishi-Lautrup fields.}
As announced, we conclude that the center-symmetric background gauge is equivalent to the Landau gauge, however, with modified boundary conditions for the fields. For instance, applying the transformation (\ref{eq:tr}) and using the symmetry (\ref{eq:sym}), we find (p.b.c. stands for periodic boundary conditions and a.b.c. for antiperiodic ones)
\beq
\frac{\int_{p.b.c.} {\cal D}a\,a^\kappa_\mu(x)\,e^{-S[\bar A_c+a,\bar A_c]}}{\int_{p.b.c.} {\cal D}[a,c,\bar c,h]\,e^{-S[\bar A_c+a,\bar A_c]}}=e^{-i\kappa \pi\tau T}\frac{\int_{p.b.c.(0)/a.b.c.(\pm)} {\cal D}a\,a^\kappa_\mu(x)\,e^{-S[a,0]}}{\int_{p.b.c.(0)/a.b.c.(\pm)} {\cal D}[a,c,\bar c,h]\,e^{-S[a,0]}},
\eeq
and, similarly,
\beq
\frac{\int_{p.b.c.} {\cal D}a\,a^\kappa_\mu(x) a^\lambda_\nu(x')\,e^{-S[\bar A_c+a,\bar A_c]}}{\int_{p.b.c.} {\cal D}[a,c,\bar c,h]\,e^{-S[\bar A_c+a,\bar A_c]}}=e^{-i\kappa \pi\tau T}e^{-i\lambda \pi\tau' T}\frac{\int_{p.b.c.(0)/a.b.c.(\pm)} {\cal D}a\,a^\kappa_\mu(x) a^\lambda_\nu(x')\,e^{-S[a,0]}}{\int_{p.b.c.(0)/a.b.c.(\pm)} {\cal D}[a,c,\bar c,h]\,e^{-S[a,0]}}.
\eeq
In particular, the two-point correlator as computed in the background Landau gauge with center-symmetric background (and with periodic boundary conditions) is related by a trivial phase factor to the two-point correlator computed in the Landau gauge with modified boundary conditions, that is, periodic ones for the color modes along $t^0$ [p.b.c.($0$)] and antiperiodic ones for the color modes along $t^\pm$ [a.b.c.($\pm$)]. This can be trivially generalized to any gauge group.

\section{Explicit calculation in the Curci-Ferrari model}\label{sec:CF}

In this section, we would like to illustrate the previous
considerations with some explicit calculations. Of course, it is not
enough to have a framework that allows one to discriminate between the
confined and deconfined phases. One also needs a good grasp on the
infrared properties that leads to the existence of a center symmetric
(confining) phase at low temperatures. In this infrared regime, the
use of the Faddeev-Popov action is inadequate due to the existence of
Gribov copies and the action (\ref{eq:action}) is expected to be
modified when these copies are properly accounted for. Here we do not
aim at taking this modification exactly. Rather, we consider a
phenomenological take on this question based on the CF
model \cite{Curci:1976bt}. In recent years, the latter has proven a powerful tool for
  an efficient perturbative description of many infrared facets of YM
  theories \cite{Tissier:2010ts,Tissier:2011ey,Pelaez:2013cpa,Gracey:2019xom,Barrios:2020ubx}, including the confinement-deconfinement transition
  \cite{Reinosa:2014ooa,Reinosa:2015gxn,Reinosa:2020mnx}. There
    is nowadays strong evidence that the main nonperturbative effects
    of YM theory can be encoded in a phenomenological parameter, the rest 
of the dynamics being weakly interacting and amenable to a perturbative treatment \cite{revue}. We
  implement the approach proposed here in this model, using the
  background-field extension of the CF Lagrangian put forward in
  Ref.~\cite{Reinosa:2014ooa}, namely,
  \beq\label{eq:S0} S[A,\bar A]=\int_x\left\{\frac{1}{4}F_{\mu\nu}^aF_{\mu\nu}^a+\bar D_\mu\bar c^a
  D_\mu c^a+ih^a\bar D_\mu(A_\mu-\bar A_\mu^a)+\frac{1}{2}m^2(A_\mu^a-\bar A_\mu^a)^2\right\}.  \eeq
The mass term in (\ref{eq:S0}) involves the aforementioned phenomenological parameter. It is tailored such that the crucial  identity (\ref{eq:sym}) is preserved and the parameter $m$ can be fixed {\it e.g.} by fitting Landau gauge correlators at zero-temperature (where the center-symmetric background gauge and the Landau gauge coincide).

We compute the effective potential $V(r,\bar r)$, obtained, up to a constant volume factor, as the effective action $\Gamma[A,\bar A]$ for field configurations of the form \eqref{eq:form} and\eqref{eq:form2}. We here briefly summarize the calculation of this potential, details will be given elsewhere. To evaluate the potential at one loop, we expand the classical action (\ref{eq:S0}) to quadratic order in the fields around a constant, temporal and diagonal gluon configuration (\ref{eq:form2}) while taking the background of the form (\ref{eq:form}). It proves convenient to switch from the usual Cartesian basis $t^a$ to the Cartan-Weyl basis $t^\kappa$ with $\kappa\in\{0,+,-\}$, introduced above. In Fourier space, the quadratic part of the action reads
\begin{equation}
\sum_\kappa \int_Q^T\,\left\{ \bar c^{-\kappa}(-Q)\,\big(\bar{Q}_\kappa\cdot {Q}_\kappa\big) \,c^\kappa(Q)+\frac{1}{2}a_\mu^{\kappa}(Q)^*\left[Q_\kappa^2P^\perp_{\mu\nu}({Q}^\kappa)+m^2\delta_{\mu\nu}\right]\,a_\nu^\kappa(Q)+h^{\kappa}(Q)^* \bar Q_\mu^\kappa a_\mu^\kappa(Q)\right\},\label{eq:quad}
\end{equation}
with $\smash{\bar Q^\kappa\cdot Q^\kappa\equiv \bar Q^\kappa_\mu Q^\kappa_\mu}$, where $\smash{\bar Q_\mu^\kappa=Q_\mu+\kappa \bar rT\delta_{\mu0}}$ and $\smash{Q_\mu^\kappa=Q_\mu+\kappa  rT\delta_{\mu0}}$. We have also introduced the notation $\int_Q^T f(Q)$ which stands for a bosonic Matsubara sum-integral $T\sum_n\int d^3q/(2\pi)^3f(\omega_n,q)$, with $\omega_n=2\pi n T$. The one-loop potential is proportional to the logarithm of the determinant of the quadratic form \eqref{eq:quad}, which can be evaluated using Schur's complement. We find
\begin{eqnarray}
 V( r,\bar r) & = & \frac{m^2T^2}{2g^2}(r-\bar r)^2+\frac{d-2}{2}\sum_\kappa\int_Q^T\ln\big[{Q}_\kappa^2+m^2\big]+\frac{1}{2}\sum_\kappa\int_Q^T\ln\left[1+\frac{m^2\bar Q_\kappa^2}{(\bar Q_\kappa\cdot{Q}_\kappa)^2}\right].\label{eq:final}
\end{eqnarray}
Since we are here interested in the dependence with respect to $r$, we can omit the contribution from $\kappa=0$. Writing $Q_\mu^\pm=\bar Q_\mu^\pm\pm(r-\bar r) T\delta_{\mu0}$, we then arrive at
\begin{eqnarray}
 V(r,\bar r) & = & \frac{m^2T^2}{2g^2}\,(r-\bar r)^2+(d-2)\int_Q^T \ln\big[ Q_+^2+m^2\big]+\int_Q^T \ln \left[1+\frac{m^2\bar Q^2_+}{(\bar Q_+^2+(r-\bar r)\bar \omega_+T)^2}\right].\label{eq:quartic}
\end{eqnarray}
We note that this expression boils down to that for $\smash{\tilde V(\bar r)\equiv V(\bar r,\bar r)}$ obtained in Ref.~\cite{Reinosa:2014ooa}, as it should. It is also straightforward to verify the symmetries discussed in Sec.~\ref{sec:constant}. A careful analysis reveals that the potential is defined only on the squares $(r,\bar r)\in[2\pi n,2\pi(n+1)]\,\times\, [2\pi n,2\pi(n+1)]$. This is no problem, however, since our proposal is to eventually choose $\smash{\bar r=\bar r_c=\pi}$ and to consider $\smash{V_c(r)=V(r,\pi)}$ within the interval 
$[0,2\pi]$ over which $V_c(r)=V_c(2\pi- r)$, thus providing a proper account 
of the center symmetry.

We note that one can absorb the center-symmetric background $\bar r_c$ into fermionic Matsubara frequencies\footnote{This corresponds to the change from periodic to antiperiodic boundary conditions for the charged color modes discussed in Sec.~\ref{sec:lattice}.} $\bar\omega_+=\omega+\pi T=(2n+1)\pi T\equiv\hat\omega$. In this case, the last sum-integral of Eq.~(\ref{eq:quartic}) becomes a standard fermionic sum-integral. The evaluation of the latter requires solving quartic equations, which is cumbersome, so we resort instead to a numerical evaluation. However, this is only possible after we properly extract the divergent part of this contribution contained in the first terms of the Taylor expansion  in powers of $r-\pi$ (which involve simpler Matsubara sums that can be performed explicitly). Disregarding once more $r$-independent contributions, we finally arrive at
\begin{eqnarray}
 V_c( r) & = & \left\{\frac{m^2}{2g^2}-\frac{1}{2\pi^2}\int_0^\infty dq q^2\,\left[\left(\frac{m^2}{q^2}+2+\frac{q^2}{m^2}\right)\frac{f_{\varepsilon_q}}{\varepsilon_q}-\left(\frac{1}{2}+\frac{q^2}{m^2}\right)\frac{f_{q}}{q}\right]\right\}(r-\pi)^2T^2\nonumber\\
& + & \frac{T}{\pi^2}\int_0^\infty dq q^2\,\ln\left(e^{-2\varepsilon_q/T}-2e^{-\varepsilon_q/T}\cos r+1\right)+\delta V_{\rm num}(r)\,,\label{eq:ff}
\end{eqnarray}
with
\beq
\delta V_{\rm num}( r) & = & \int_{\hat Q}^T \left[\ln \left(1+\frac{m^2\hat Q^2}{(\hat Q^2+(r-\pi)\hat\omega T)^2}\right)-\ln\left(1+\frac{m^2}{\hat Q^2}\right)-(r-\pi)^2T^2\frac{\hat\omega^2(3\hat Q^2+m^2)}{\hat Q^4(\hat Q^2+m^2)^2}\right].\label{eq:22}
\eeq
Here, $\int_{\hat Q}^T$ stands for a fermionic Matsubara sum  $T\sum_n\int d^3q/(2\pi)^3f(\omega_n,q)$,with $\omega_n=2\pi (n+1/2) T$ and $f_\varepsilon=1/(e^{\beta\varepsilon}+1)$ is the Fermi-Dirac distribution. The latter appears due to the presence of the confining (center-symmetric) background, which changes the boundary conditions of those fields with $\kappa=\pm$ and thus the thermal distribution from Bose-Einstein to negative Fermi-Dirac: $n_{\varepsilon\pm i\pi T}=-f_{\varepsilon}$.
In writing (\ref{eq:ff}) we have absorbed all zero-temperature contributions proportionnal to $(r-\pi)^2$ in the renormalization of the tree-level term. In other words, our renormalization scheme is such that, 
the curvature of the potential in the $r$-direction at $r=\pi$, 
equals $m^2/g^2$ at zero temperature. We have checked that the counterterms needed are just the usual counterterms at zero temperature, in the so-called vanishing momentum scheme for the propagators \cite{Tissier:2011ey}. In our numerical application below, we shall then use the values of the parameters obtained from fitting zero-temperature propagators in this scheme to the lattice data (in the Landau gauge). Again, details will be given elsewhere.

\begin{figure}[h]
  \centering
  \includegraphics[width=0.45\linewidth]{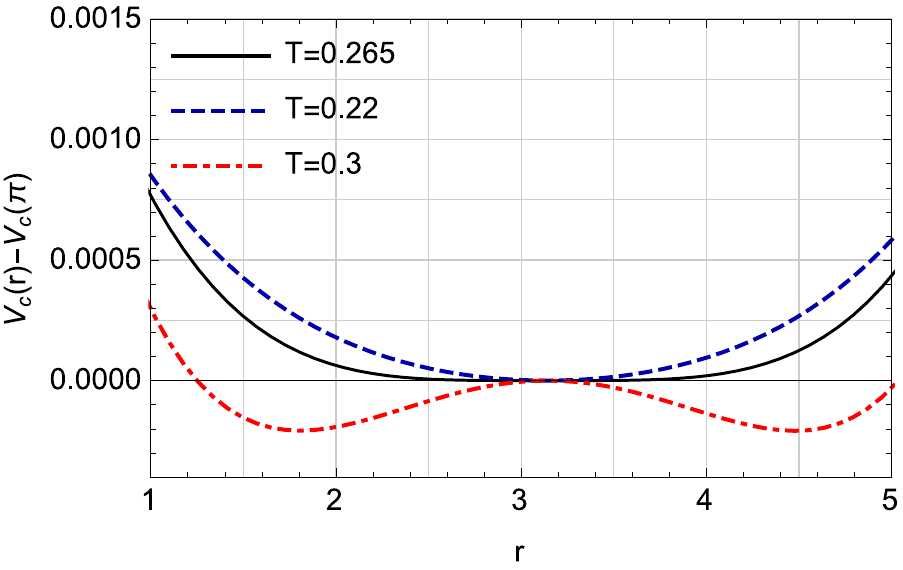}\qquad   \includegraphics[width=0.45\linewidth]{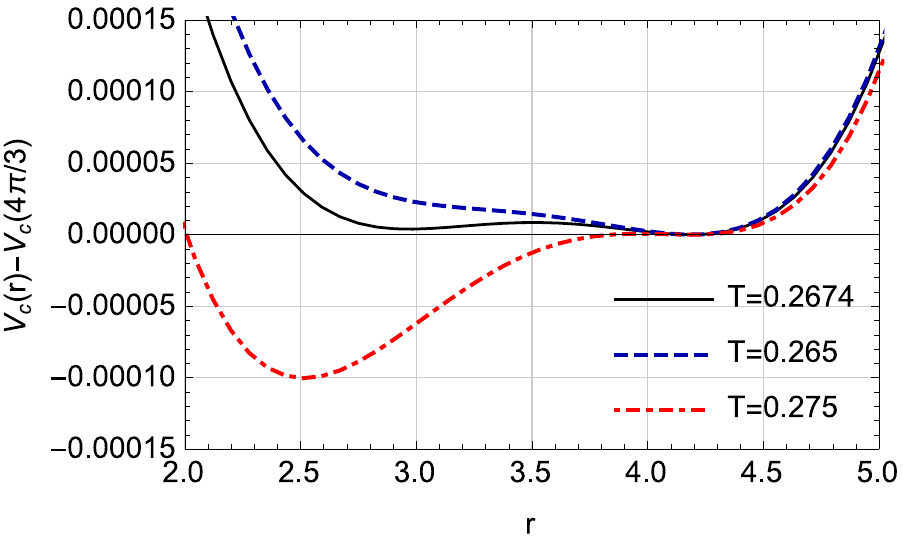}
 \caption{The potential $V_c(r)$ in the SU(2) case (left) and in the SU(3) case (right).}\label{fig:su2_pot}
\end{figure}

In Fig.~\ref{fig:su2_pot}, we show the behavior of the potential $V_c(r)$ as a function of the temperature. We clearly observe a continous transition characteristic of the SU(2) case. The value of $T_c$ can be determined from solving $V_c''(\pi)=0$. Since $\delta V_{\rm num}''(\pi)=0$ by construction, one easily deduces that
\begin{eqnarray}\label{eq:Debye}
V_c''(\pi) = \frac{m^2}{g^2}-\frac{1}{2\pi^2}\int_0^\infty dq\,q^2\left[\left(6\frac{m^2}{q^2}+12+2\frac{q^2}{m^2}\right)\frac{f_{\varepsilon_q}}{\varepsilon_q}-\left(1+2\frac{q^2}{m^2}\right)\frac{f_q}{q}\right].
\end{eqnarray}
As compared to the determination of the transition temperature using $\smash{\tilde V''(\pi)=0}$, see Ref.~\cite{Reinosa:2014ooa}, we here need to provide both $m$ and $g$. With the parameters $m=680$~MeV and $g=7.5$ obtained by fitting one-loop zero-temperature CF propagators to lattice data (in the same zero-temperature renormalization scheme as the one used here) \cite{Tissier:2011ey}, we find $ T_c\simeq 265$~MeV to be compared to the result $\bar T_c\simeq 227$~MeV obtained with the same value of $m$ and $\tilde V(\bar r)$, and more importantly to the lattice value of $295$~MeV \cite{Lucini:2012gg}.\footnote{In Ref.~\cite{Reinosa:2014ooa}, $\bar T_c\simeq 237$~MeV was determined using the value $m=710$~MeV obained by fitting tree-level CF propagators to the lattice data.}

\begin{figure}[t]
  \centering
  \includegraphics[width=0.45\linewidth]{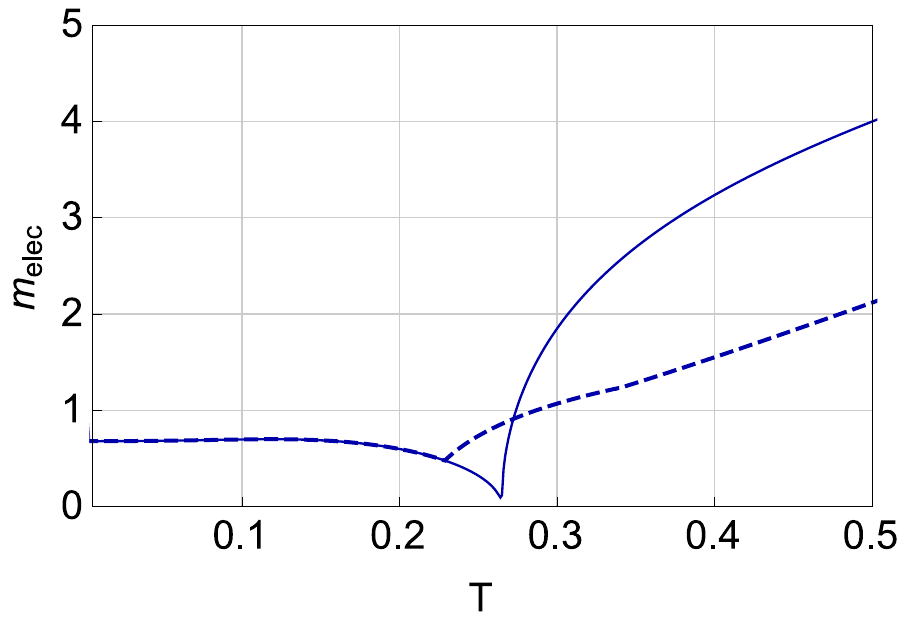}\qquad\includegraphics[width=0.45\linewidth]{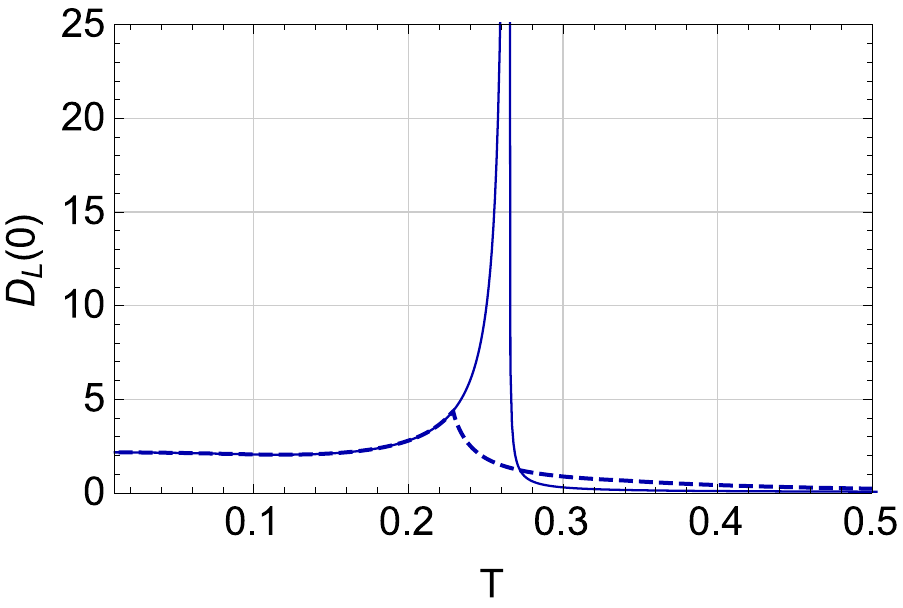}\\
    \includegraphics[width=0.45\linewidth]{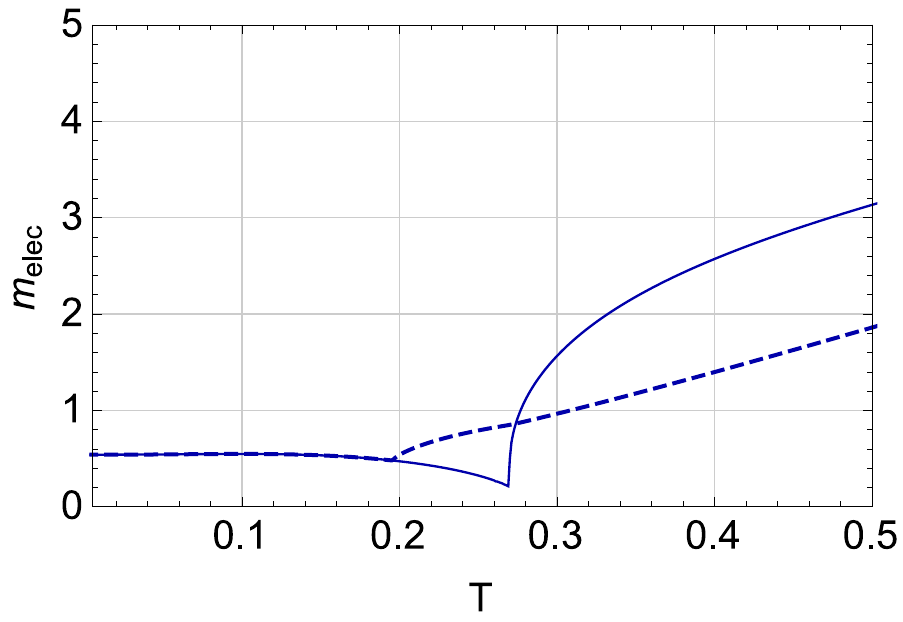}\qquad\includegraphics[width=0.45\linewidth]{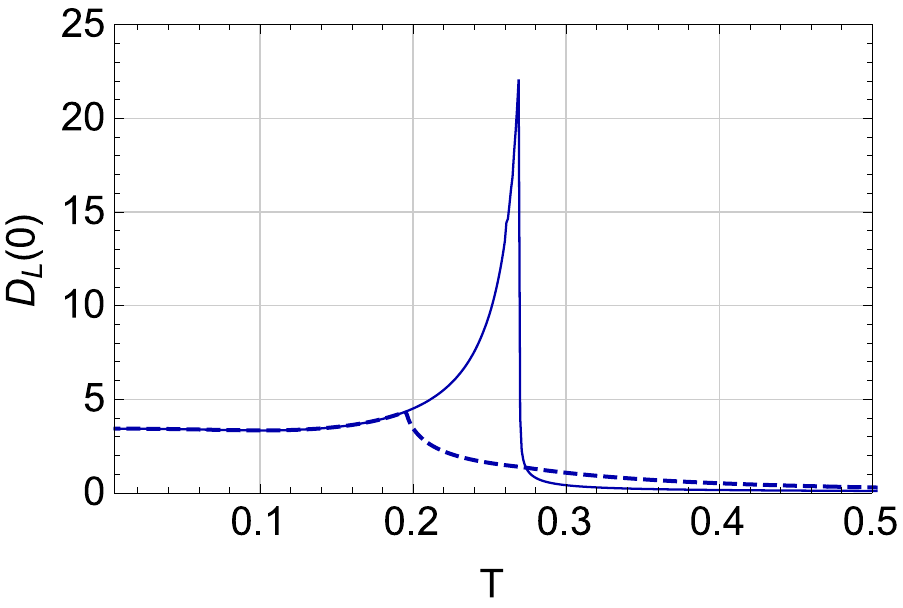}
 \caption{The neutral electric mass (left) and its inverse, the neutral electric susceptibility or longitudinal propagator $D_L(0)$ at vanishing frequency and momentum (right) as functions of the temperature for the SU($2$) theory (upper pannel) and for the SU($3$) theory (lower pannel). The dashed lines show the corresponding results obtained in the self-consistent background approach \cite{Reinosa:2016iml}. $T$ and $m$ are given in GeV.}\label{fig:two}
\end{figure}

The inverse curvature of the potential at its minimum, $1/V_c''(r_{\rm min})$ is nothing but the electric gluon propagator for the neutral color mode at vanishing momentum and frequency, also called the (neutral) electric susceptibility. Here, for convenience, we shall call the curvature $V_c''(r_{\rm min})$ itself the neutral electric square mass. Because this quantity is gauge variant, it does not make sense {\it a priori}  to compare it in different gauges. However, the determinations of the electric mass in the center-symmetric and in the self-consistent background gauges should coincide over the whole confining phase and vanish when approaching the transition. This is because both backgrounds coincide in that phase and we check explicitly that coincidence at low temperatures, see Fig.~\ref{fig:two}. However, due to the approximations, the very determination of the upper end of the confining phase depends on the gauge. In fact, at the present order, we have $\bar T_c<T_c$ and the electric mass in the self-consistent background gauge never reaches zero. Instead, in the center-symmetric background gauge, $T_c$ is directly determined from the condition $V''_c(\pi)=0$ and the neutral electric mass vanishes at the transition $T_c$ by construction (at all orders). This is shown in Fig.~\ref{fig:two} together with the behavior of the neutral electric susceptibility. \\

The discussion can be  easily extended to the SU(3) case. In fact the formula for the potential (\ref{eq:final}) remains the same provided one replaces the color labels $\kappa$ by those of the corresponding Cartan-Weyl basis. In the SU(3) case, the labels $\kappa$ become two-dimensional vectors, with two degenerate {\it zeros} and six {\it roots} $\alpha=(\pm 1,0)$, $(1/2,\pm\sqrt{3}/2)$, $(-1/2,\pm\sqrt{3}/2)$. The background $\bar r$ and the field $r$ are also two dimensional vectors with components along the diagonal directions $3$ and $8$ of the algebra and $\bar Q^\kappa_\mu$ and $ Q^\kappa_\mu$ should now be understood as $\bar Q^\kappa_\mu= Q_\mu+(\kappa\cdot \bar r)\, T\delta_{\mu0}$ and $Q^\kappa_\mu= Q_\mu+(\kappa\cdot r)\,T\delta_{\mu0}$ respectively. Since YM theory is charge-conjugation invariant, we can, without loss of generality, restrict to $r_8=0$ in which case only the first component of the roots matters that is $\pm 1$ and (twice) $\pm 1/2$. It follows that, if we denote by $\Delta V_{\mbox{\tiny SU(2)}}(r,\bar r)$ the last two terms in Eq.~(\ref{eq:quartic}), the relevant SU(3) potential writes ($\bar r_c= 4\pi/3$ and $ r\equiv  r_3$)
\beq
V_c^{\mbox{\tiny SU(3)}}(r_3)=\frac{m^2T^2}{2g^2}\left(r_3-\frac{4\pi}{3}\right)^2+\Delta V_{\mbox{\tiny SU(2)}}\left(r_3,{4\pi\over3}\right)+2\Delta V_{\mbox{\tiny SU(2)}}\left( {r_3\over2},{2\pi\over3}\right).\label{eq:su3_pot}
\eeq
The potential is shown in Fig.~\ref{fig:su2_pot}. With the parameters $m=540$~MeV and $g=4.9$, obtained (in the same scheme) by fitting the one-loop zero-temperature CF propagators to the lattice data in the Landau gauge \cite{Tissier:2011ey}, we find a first order phase transition at $ T_c\simeq 267$~MeV, rather close to the value of $270$~MeV obtained in lattice simulations \cite{Lucini:2012gg}. The estimate using the background potential $\tilde V(\bar r)$ and the same value of the mass gives instead a transition temperature at $\bar T_c\simeq 197$~MeV.\footnote{In Ref.~\cite{Reinosa:2014ooa}, $T_c\simeq 185$~MeV was determined using the value $m=510$~MeV obained by fitting tree-level CF propagators to the lattice data.} We note that the agreement with the lattice value is better in the SU(3) case than in the SU(2) case. This is in line with the fact that the perturbative expansion in the CF model is more under control in the SU(3) case due to the fact that the coupling is smaller \cite{Tissier:2011ey}. As for the neutral electric mass, it is considerably reduced when approaching the transition from below but never vanishes. This corresponds to a sharp peak in the susceptibility rather than a singularity as in the SU(2) case, see Fig.~\ref{fig:two}. We mention that, although this is not really visible in the figure, the SU(3) neutral electric mass is discontinuous at the transition, with $m_{\rm elec}(T_c^-)>m_{\rm elec}(T_c^+)$. Correspondingly, the peak in the susceptibility is also (slightly) discontinous.

\begin{figure}[t]
  \centering
  \includegraphics[width=0.4\linewidth]{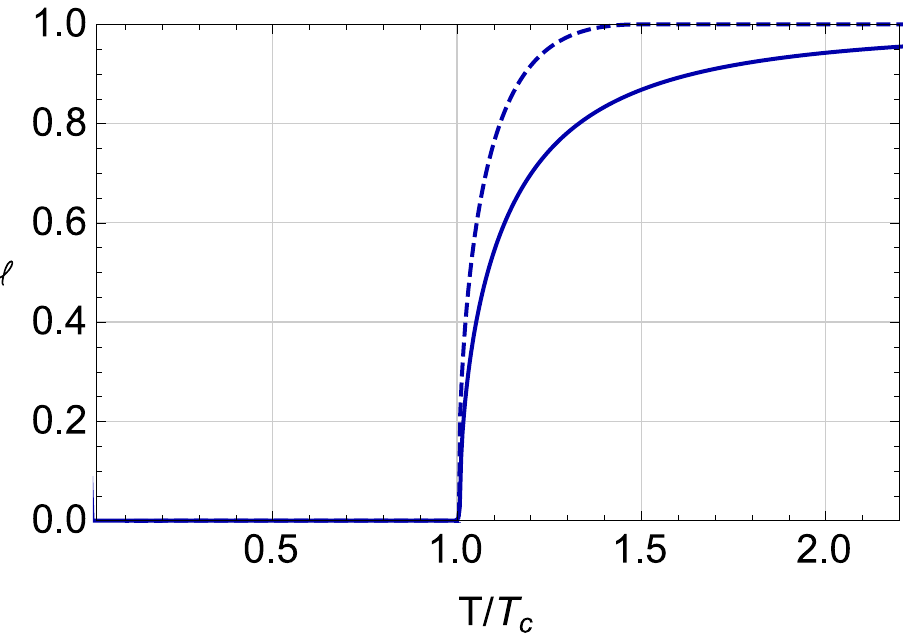}\qquad\qquad\includegraphics[width=0.4\linewidth]{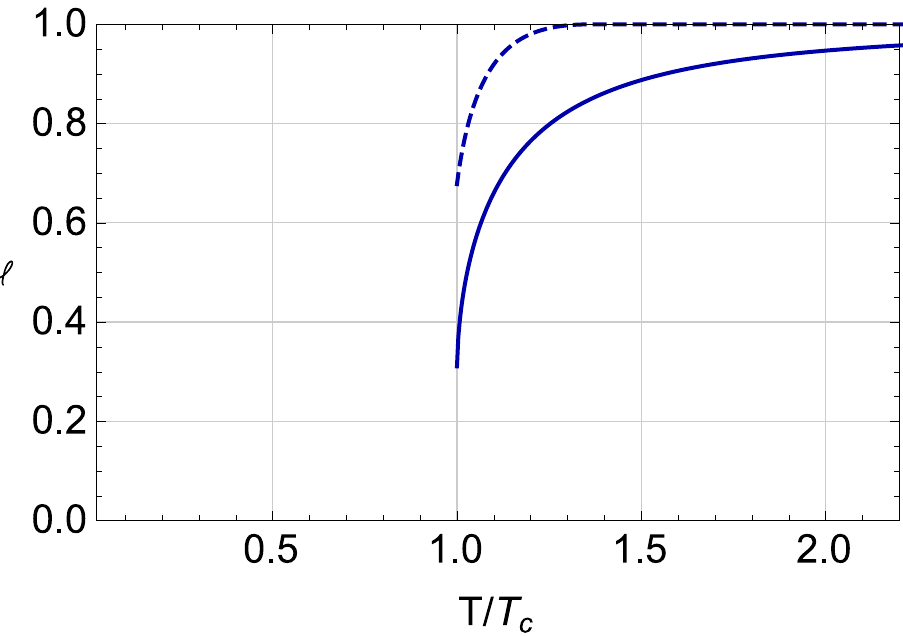}
 \caption{SU(2) (left) and SU(3) (right) Polyakov loops as obtained using the self-consistent background field approach \cite{Reinosa:2014ooa} (dahed) or the present center-symmetric background approach (plain).}\label{fig:loops0}
\end{figure}

\begin{figure}[t]
  \centering
  \includegraphics[width=0.42\linewidth]{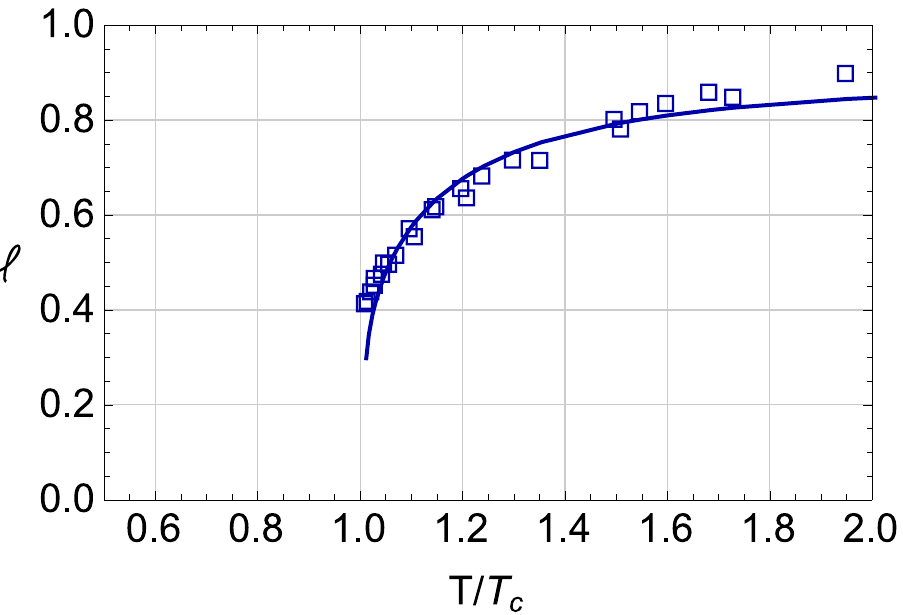}
 \caption{Comparison of our SU(3) Polyakov to the lattice data of Refs.~\cite{Gupta:2007ax,Lo:2013hla}.}\label{fig:loops}
\end{figure}

Finally, we compute the Polyakov loop (\ref{eq:ell}). Following, for instance, Ref.~\cite{Reinosa:2014zta}, we introduce a function $\ell(r,\bar r)$ that gives the Polyakov loop when $r$ is evaluated at the minimum $r_{\rm min}(\bar r)$ of the potential. At the order considered here, we have, formally,
\beq
\ell(r,\bar r)=\ell_0(r,\bar r)+g^2 \ell_1(r,\bar r)+{\cal O}(g^4)\,.
\eeq
Center-symmetry implies that $\ell(\pi,\pi)=0$ at all orders and, thus, $\ell_1(\pi,\pi)=0$. Because, in the  present gauge, $\bar r=\pi$ and $r_{\rm min}\equiv r_{\rm min}(\pi)=\pi+{\cal O}(g^2)$, we conclude that the next-to-leading order Polyakov loop is given by its tree-level expression $\ell_0(r_{\rm min},\pi)$, that is,
\beq
\ell=\frac{1}{N} {\rm tr}\,e^{i r^j_{\rm min} t^j}+{\cal O}(g^4)\,,
\eeq
where $r_{\rm min} $ is to be evaluated from the next-to-leading-order potential \eqref{eq:final}. This is yet another strength of the approach proposed here. The result is shown in Fig.~\ref{fig:loops0} in comparison with the results previously obtained using the self-consistent background field approach \cite{Reinosa:2014ooa}. We observe that the rise of the Polyakov loop in the deconfined phase towards its maximal value is systematically smaller when using $V_c(r)$, which is in line with what is observed in lattice simulations. A comparison with lattice data requires to take into account the renormalization of the Polyakov loop. Here, we have computed the bare Polyakov in dimensional regularization which should be seen as the renormalized Polyakov loop in a certain scheme. Since this is not the same scheme as the one on the lattice, we need to allow for an overall (temperature-independent) normalization. In the SU(3) case (where the perturbative CF approach works best), we find that the best fit is obtained over the interval $[T_c,2T_c]$, see Fig.~\ref{fig:loops}. Above $2T_c$, our result departs from the lattice one. This is expected as various effects not included here, such as the resummation of hard thermal loops \cite{Haque:2014rua} or renormalisation group running \cite{Herbst:2015ona,Kneur:2021feo}, are known to play an important role in this regime. Needless to mention, this is just a very simple one-loop calculation. Taking into account higher order corrections as well as the above mentioned effects is work is under way.

\section{Conclusions}
We have investigated a new approach within background field gauges at finite temperature. It is based on the use of the effective action $\smash{\Gamma[A,\bar A=\bar A_c]}$ with a fixed center-invariant background $\bar A_c$ rather than the self-consistent background field effective action $\smash{\tilde\Gamma[\bar A]=\Gamma[A=\bar A,\bar A]}$. In principle, the two approaches should be equivalent in the confined phase. However, the equivalence relies on a property that is generally violated in continuum approaches, namely, the exact invariance of the gauge-fixed partition function on the background field $\bar A$. Our new proposal does not rely on this property and, as such, should provide more reliable results. In particular it allows to directly assess the impact of the deconfinement transition on the gluon propagator and implies that the zero-momentum/frequency value of the latter diverges at the transition in the SU(2) case. We have argued that this prediction is easily testable within lattice simulation since the center-symmetric gauge proposed here can be implemented with current techniques used for the Landau gauge through a twist of the boundary conditions for the gauge fields.

We have also investigated these predictions within the CF model. We find indeed that the SU(2) neutral electric mass vanishes at the transition. In addition, we obtain new predictions for the transition temperature. Quite remarkably, in the SU(3) case (where the pertubative CF model works best) the predicted value at one-loop order is already pretty close to the lattice result. Also interesting is the fact that the rise of the Polyakov in the deconfined phase is slower than the one previously obtained using the self-consistent background field approach. This is in line with what is observed on the lattice (for the renomalized Polyakov loop) \cite{Kaczmarek:2002mc,Gupta:2007ax,Lo:2013hla}. We are currently investigating the full momentum dependence of the propagators across the transition in this novel approach. It will be also interesting to evaluate the impact of two-loop corrections on the present results.

As a final word, let us mention that it is tempting to draw similarities between our results in the SU(2) case and lattice results for the electric susceptibility in the Landau gauge \cite{Cucchieri11,Cucchieri12}. One could in fact wonder whether the gauge propagator in the presence of a center-invariant background  can teach us something about the Landau gauge propagator. There are two main reasons why we could envisage this possibility.

The first one lies in the similarities between the perturbatives expressions for the susceptibilities along the diagonal color direction in the Landau and in the Landau-DeWitt gauges. Indeed these two quantities are given by the same perturbative diagrams, with the only difference that, in the latter case, the internal frequencies associated to the color charge of the internal lines are shifted by a quantity proportional to the background field \cite{Reinosa:2015gxn,Reinosa:2020mnx}. This pretty much resembles the relation between the pressure in the Landau and in the Landau-DeWitt gauges. For this latter quantity, since the pressure is gauge-independent, there should exist a mechanism explaining how one can move from an infinite set of diagrams with shifted internal frequencies, to the same diagrams with no shifted frequencies. If the same mechanism holds for the susceptibility along the diagonal color direction, this could open a connection with the same quantity computed in the Landau gauge.\footnote{This would not mean that this quantity is gauge-independent, just that it coincides in two different gauges.}

A second argument in favor of a connection between the propagators in the Landau and the Landau-DeWitt gauges is that our argumentation leads to the conclusion that the background field propagator (with center-invariant background) has a pole at vanishing momentum for $\smash{T=T_c}$. Poles of YM correlations functions are expected to be gauge-independent, so the Landau gauge propagator should also feature a pole at vanishing momentum for $\smash{T=T_c}$. This assumes, however, that the gauge-independence of poles extends to background gauges, which is not clear {\it a priori} \cite{Kobes:1990dc}. Further investigations in this direction would require studying the background Nielsen identities.

\end{document}